\documentclass[amsmath,twocolumn,superscriptaddress,showpacs,aps,prb]{revtex4}
\usepackage{bm}
\usepackage{graphics}
\begin{document}

\title{Transport equations for a two-dimensional electron gas with spin-orbit interaction}
\author{E.G. Mishchenko}
 \affiliation{Lyman
Laboratory, Department of Physics, Harvard University, MA 02138}
\author{B.I. Halperin}
\affiliation{Lyman Laboratory, Department of Physics, Harvard
University, MA 02138}

\begin{abstract}
The transport equations for a two-dimensional electron gas with
spin-orbit interaction are presented. The distribution function is
a $2\times 2$-matrix in the spin space. Particle and energy
conservation laws determine the expressions for the electric
current and the energy flow. The derived transport equations are
applied to the spin-splitting of a wave packet and to the
calculation of the structure factor and the dynamic conductivity.
\end{abstract}


 \maketitle Spin injection and coherent control of spins in
various nanostructures represent two principal challenges for the
field of spintronics. Recently, the amount of spintronics research
has grown up extensively with the ultimate goal of applications to
the quantum computing and information processing \cite{ALS}. A
number of spin-based devices have been designed and studied
\cite{DD,WAB,KNT,GBZ}. Spin manipulation in such devices can be
achieved by optical \cite{KA} or electric \cite{FKR,OYB,ZRK,RPS}
methods or by ferromagnetic gating \cite{fer}. A controlled
coupling between spin and orbital degrees of freedom is considered
to be a particularly promising tool of efficient spin manipulation
dating back to the seminal proposal by Datta and Das \cite{DD}.

Spin-orbit interaction in two-dimensional electron gas (2DEG)
confined at GaAs/AlGaAs, GaN/AlGaN or similar heterojunctions
arises because of the quantum well asymmetry in the perpendicular
[$z$] direction. The resulting perpendicular electric field leads
to the coupling of spin to the electron momentum \cite{BR}. The
strength of this coupling can be experimentally tuned by a gate
voltage \cite{NAT,Koga}.

Experimental advances in spin manipulation present a certain
challenge to develop a proper theoretical description for various
phenomena related to the spin-orbit interaction. In particular,
modification of universal conductance fluctuations and weak
localization has been studied in quantum dots \cite{H,AF,M}. The
phenomenon of weak localization has been considered in 2DEG as
well \cite{Koga,MZM}. The Friedel oscillations in the presence of
spin-orbit interaction\cite{CR} and the ac conductivity and the
plasmon attenuation\cite{MCE}  are calculated.

The principal goal of the present paper is to derive general
transport  equations for the spin-dependent distribution function
of 2DEG including the effects of spin-orbital coupling. We assume
that the spin-orbit interaction in a two-dimensional electron gas
has the form,
\begin{equation}
\label{soham}
 H_{so} =\alpha(\hat\sigma_x p_y-\hat\sigma_y p_x)
+\beta(\hat\sigma_x p_x-\hat\sigma_y p_y),
\end{equation}
where the first term is the Bychkov-Rashba term \cite{BR}  and the
second term is the linear Dresselhaus (or anisotropy) term present
in semiconductors with no bulk inversion symmetry \cite{D}. The
expression of Eq.\  (\ref{soham}) corresponds to the confinement
along the (001) growth direction \cite{DK}.  Hereinafter we
neglect cubic Dresselhaus terms. The free particle Hamiltonian can
therefore be written in compact notations as,
\begin{equation}
\label{ham}
H= [\frac{{\bf p}^2}{2m}-\mu]\hat\sigma_0+ \alpha_{ik} \hat\sigma_i p_k, ~~ 
~\alpha_{ik} =\left(
\begin{array}{cc}  \beta & \alpha \\ -\alpha & -\beta \end{array}
\right).
\end{equation}
Here, as usual, $\hat\sigma_0, \hat\sigma_x,\hat\sigma_y,
\hat\sigma_z$ constitute the set of Pauli matrices. We use Latin
subscripts for spatial coordinates and reserve Greek subscripts
for the spin indexes.

{\it Spectral properties.} Before deriving the transport equation
we describe briefly the spectral properties of the Hamiltonian
(\ref{soham}). Its diagonalization is straightforward and reveals
the existence of  two spin-split subbands in the electron
spectrum,
\begin{equation}
\label{spectrum} \epsilon_{1{\bf p}} = \xi_p+\Delta_{\bf p},~~~
\epsilon_{2{\bf p}} =\xi_p-\Delta_{\bf p},
\end{equation}
where
\begin{equation}
\label{spectrum1} \xi_p=\frac{p^2}{2m}-\mu, ~~~ \Delta_{\bf
p}=\sqrt{p^2(\alpha^2+\beta^2)+4\alpha\beta p_xp_y },
\end{equation}
and $p^2=p_x^2+p_y^2$ is the total electron momentum. The
eigenfunctions corresponding to the eigenstates (\ref{spectrum})
are,
\begin{equation}
\label{states} \psi_{1,2} ({\bf x}) =\frac{1}{\sqrt{2}} \left(
\begin{array}{c} e^{i\chi_{\bf p}/2}
\\ \pm e^{-i\chi_{\bf p}/2}
 \end{array} \right) e^{i{\bf p x}},
\end{equation}
with the upper (lower) sign corresponding to the  $\psi_1$
($\psi_2$) state. The phase factor $\chi_{\bf p}$ depends on the
direction of the electron momentum,
\begin{equation}
\tan \chi_{\bf p} = \frac{\alpha p_x+\beta p_y}{\alpha p_y +\beta
p_x}.
\end{equation}
For the isotropic 2DEG ($\beta =0$) the phase $\chi_{\bf p}$
coincides with the angle between the electron momentum and the
$y$-axis.

Further and more convenient description of the spectral properties
can be obtained by considering the spin-dependent retarded Green
function, defined as usual by,
\begin{equation}
\label{gr} i G^R_{\alpha \beta}(x,x') =\theta(t-t') \langle
\langle \psi_\alpha (x)\psi_\beta^\dagger (x')+\psi_\beta^\dagger
(x') \psi_\alpha (x) \rangle \rangle.
\end{equation}
Here we have used the shorthand notation for the space and time
variables, $x=({\bf x},t)$. In a homogeneous system the
correlation functions depend on the relative coordinates $x-x'$
only. Using the above expressions (\ref{spectrum}-\ref{states}) we
can write the expression for the retarded Green function  of free
electrons in the momentum representation, $\hat G^R (\epsilon,{\bf
p}) =\int dt d{\bf x} ~\hat G^R (x) e^{i\epsilon t-i{\bf p}{\bf
x}}$, which after simple transformations takes the form,
\begin{eqnarray}
\label{greq} \hat G^R(\epsilon,{\bf p})&=& \sum_{\mu=1,2}\hat
G_\mu^R(\epsilon,{\bf p}) = \frac{1}{2} \frac{\hat
\sigma_0+\cos{\chi_{\bf p}}\hat\sigma_x
-\sin{\chi_{\bf p}}\hat \sigma_y}{\epsilon -\epsilon_{1{\bf p}}+i\eta}\nonumber\\
&& +\frac{1}{2} \frac{\hat \sigma_0-\cos{\chi_{\bf p}}\hat\sigma_x
+\sin{\chi_{\bf p}}\hat \sigma_y}{\epsilon -\epsilon_{2{\bf
p}}+i\eta},
\end{eqnarray}
with the indexes $\mu=1,2$ corresponding to the first and second
terms, respectively, in the last expression of Eq.\ (\ref{greq}).

The central quantity in the transport theory is the density
matrix,
\begin{equation}
\label{densitym} f_{\alpha \beta}(x,x') = \langle \langle
\psi_\beta^\dagger (x') \psi_\alpha (x) \rangle \rangle.
\end{equation}
Its value in the thermal equilibrium is related to the imaginary
part of the retarded Green function via the
fluctuation-dissipation theorem,
\begin{equation}
\label{fdt} \hat f(\epsilon,{\bf p}) = \sum_{\mu=1,2} n_{\mu\bf p}
~[ \hat G^{R\dagger}_\mu (\epsilon,{\bf p})-G^{R}_\mu
(\epsilon,{\bf p})],
\end{equation}
here $n_{\mu \bf p}$ is the Fermi-Dirac distribution function for
the $\mu$-th state (\ref{spectrum}). It is convenient to expand
the density matrix over the complete set of the Pauli matrices. In
particular, for the equal-time density matrix,
\begin{equation}
\label{pauli} \hat f_{\bf p} = \int \frac{d\epsilon}{2\pi} \hat
f(\epsilon,{\bf p})= \frac{1}{2} f_{\bf p}\hat\sigma_0+\frac{1}{2}
{\bf g}_{\bf p}\cdot \hat{\bm{\sigma}},
\end{equation}
we observe according to Eq.\ (\ref{fdt}) that in the thermal
equilibrium,
\begin{equation}
\label{equil}
\begin{array}{ll} f_{\bf p}=(n_{1\bf p}+n_{2\bf p}),&  ~g_{{\bf p}x}=\cos\chi_{\bf p}
(n_{1\bf p}-n_{2\bf p}),\\ g_{{\bf p}z}=0, & ~g_{{\bf
p}y}=-\sin\chi_{\bf p}(n_{1\bf p}-n_{2\bf p}),
\end{array}
\end{equation}

{\it Transport equations}. In a generic nonequilibrium state, the
density matrix (\ref{densitym}) obeys a set of conjugated
equations that can be obtained from the equations of motion for
the electron operators $\psi(x)$ and $\psi^\dagger(x)$ determined
by the Hamiltonian (\ref{ham}),
\begin{eqnarray}
\label{set} \left[ i\partial_t +\frac{\nabla^2}{2m}+\mu
-e\phi_x\right]\hat f(x,x')
+i\alpha_{ik} \hat\sigma_i \nabla_k \hat f(x,x')=0,\nonumber\\
\left[i\partial_{t'} -\frac{\nabla'^2}{2m}-\mu
+e\phi_{x'}\right]\hat f(x,x') +i\alpha_{ik} \nabla_k' \hat
f(x,x')\hat \sigma_i =0.\nonumber\\
\end{eqnarray}
The equations (\ref{set}) neglect impurity scattering. This is
justified for ballistic systems  when the mean free path exceeds
the characteristic system size, e.g.\ in high-mobility 2DEG in
semiconductor heterostructures (we discuss the impurity scattering
at the end of the paper). In Eq.\ (\ref{set}) we allowed for the
scalar external field $\phi_x=\phi({\bf x},t)$. In the absence of
electron-impurity or electron-electron collisions no self-energy
terms appear in the equations (\ref{set}) which makes it
sufficient to consider the equal-time ($t=t'$) functions only.

Following the known route of deriving kinetic equations \cite{LP}
we utilize the  Wigner transformation for the density matrix,
\begin{equation}
\label{wign} \hat f_{\bf p}({\bf x}, t)= \int d {\bf r}~ e^{-i \bf
pr} \hat f({\bf x}+\frac{\bf r}{2}, {\bf x}-\frac{\bf r}{2},t ).
\end{equation}
By taking the sum of equations (\ref{set})  in the Wigner
representation we obtain,
\begin{eqnarray}
\label{tr} [ \partial_t +{\bf v}\cdot\nabla ] \hat f_{\bf
p}+ie\int d{\bf q}~\phi_{\bf q}
(\hat f_{{\bf p}-\frac{\bf q}{2}}-\hat f_{{\bf p}+\frac{\bf q}{2}})e^{i{\bf qx}}\nonumber \\
+ i\alpha_{ik} p_k[\hat\sigma_i,\hat f_{\bf p}]+\frac{1}{2}
\alpha_{ik}\nabla_k \{\hat\sigma_i, \hat f_{\bf p}\}=0,
\end{eqnarray}
where ${\bf v}={\bf p}/m$. Here we introduced the spatial Fourier
transform for the scalar potential $\phi_{\bf q}= \int d{\bf x}
\phi({\bf x},t)e^{i{\bf qx}}$, with the shorthand notation for the
momentum integration $d{\bf q}=d^2q/(2\pi)^2$.

Finally, to present Eq. (\ref{tr}) in a more transparent way we
turn to the Pauli matrix representation (\ref{pauli}) to write,
\begin{eqnarray}
\label{tr1} [ \partial_t +{\bf v}\cdot\nabla]~ f_{\bf p}&+&ie\int
d{\bf q}~\phi_{\bf q} (f_{{\bf p}-\frac{\bf q}{2}}-f_{{\bf
p}+\frac{\bf q}{2}})e^{i{\bf qx}}\nonumber \\ &+& \alpha_{ik}
\nabla_k g_{{ \bf p}i} =0,
\end{eqnarray}
\begin{eqnarray}
\label{tr2} [ \partial_t +{\bf v}\cdot\nabla] ~ g_{{\bf p }i}
&+&ie\int d{\bf q}~\phi_{\bf q} (g_{{\bf p}-\frac{\bf
q}{2}i}-g_{{\bf p}+\frac{\bf q}{2}i})e^{i{\bf qx}}\nonumber\\
\nonumber\\ &-&[{\bf b}_{\bf p}\times {\bf g}_{\bf p}]_i
+\alpha_{ik} \nabla_k f_{\bf p}=0,
\end{eqnarray}
with the following notation for the precession frequency, ${ b}_{
{\bf p}i}= 2\alpha_{ik}p_k = 2\Delta_{\bf p} (\cos\chi_{\bf
p},-\sin{\chi_{\bf p}},0)$.

{\it Conservation laws.} The transport equations
(\ref{tr1},\ref{tr2}) are of the Boltzmann type and therefore
fulfill certain particle and energy conservation conditions which
will now be obtained. By integrating Eq. (\ref{tr1}) with respect
to the momentum we find the continuity equation for the particle
flow,
\begin{equation}
\label{cont}  \partial_t \rho +\frac{1}{e}\nabla \cdot {\bf j} =0,
\end{equation}
where the electron density and the electric current are given
respectively by,
\begin{eqnarray}
\label{density}
 \rho = \int d{\bf p} f_{\bf p},~~~ j_k = e\int
d{\bf p}~ [v_k f_{\bf p}+ \alpha_{ik} g_{{\bf p}i}].
\end{eqnarray}
The terms containing the external potential $\phi_{\bf q}$ cancel
as is readily seen by the change of integration variables. To
obtain the energy continuity condition we multiply Eq. (\ref{tr1})
by $\xi_p$ and Eq. (\ref{tr2}) by ${\bf b}_{\bf p}$ and add them
together. After simple transformations the conservation of energy
can be written in the conventional form,
\begin{equation}
\label{heat}
\partial_t \rho^\epsilon +\nabla \cdot {\bf j}^\epsilon ={\bf j}\cdot {\bf E},
\end{equation}
where the energy density and energy current are,
\begin{eqnarray*}
\rho^\epsilon &=& \int d{\bf p} [\xi_p f_{\bf p}+ {b}_{{\bf
p}i} {g}_{{\bf p}i}],\nonumber\\
j^\epsilon_k &=& \int d{\bf p} v_k[\xi_p f_{\bf p}+ {b}_{{\bf
p}i}g_{{\bf p}i}]+\alpha_{ik} \int d{\bf p}[\xi_p g_{{\bf p}i}
+b_{{\bf p}i} f_{\bf p}].
\end{eqnarray*}
The equation (\ref{heat}) means that the local energy change is
due to the energy flow to the neighboring points in space as well
as a result of  the local Joule heating (right-hand side).

{\it Wave packet splitting.} To give a specific application of the
derived equations let us now use them to describe the propagation
of a wave packet in 2DEG with a spin-orbit coupling. We neglect a
spin-orbit anisotropy $\beta=0$ for simplicity. The wave packet
propagates along the $y$-direction and is uniform along the
$x$-axis. The transport equations (\ref{tr1},\ref{tr2}) are then
one-dimensional and (with no external field applied) yield,
\begin{eqnarray}
\label{packet}
&&\left[ \partial_t +v\partial_y \right] f=-\alpha \partial_y g_x,\nonumber\\
&& \left[ \partial_t + v\partial_y\right]g_x=-\alpha \partial_y
f,\nonumber\\
&&\left[ \partial_t + v\partial_y\right]g_y=-2\Delta_p g_z
\nonumber\\
&& \left[
\partial_t + v\partial_y\right]g_z= 2\Delta_p g_y.
\end{eqnarray}
First, we consider a spin-unpolarized Gaussian wave packet
injected at the point $y=0$ at the time $t=0$ and moving with the
average momentum $\bar p$,
\begin{equation}
\label{fp} \hat f_{\bf p}({\bf x},t=0) = \hat \sigma_0 F (y), ~~
F(y)=e^{-y^2\delta p^2 -\frac{(p_y-\bar p)^2}{\delta p^2}}.
\end{equation}
In this geometry the phase factor $\chi_{\bf p}=0$, which means
that the precession vector ${\bf b}$ is directed along the
$x$-axis. We also observe that $g_y=g_z=0$.  The remaining two of
the equations (\ref{packet}) are easily solved by Fourier
transforming them into a set of linear algebraic equations.
The general solution of Eqs. (\ref{packet}) takes the form,
\begin{eqnarray*}
f(y,t) = A(y-v_+t)+B(y-v_-t),\\
g_x(y,t) = A(y-v_+t)-B(y-v_-t),
\end{eqnarray*}
where we have introduced subband velocities $v_\pm = v \pm
\alpha$. So far, $A(x)$ and $B(x)$ are two arbitrary functions
which have to be determined from the initial condition
(\ref{packet}) yielding, $ A(y)=B(y) = F(y) $. We find that the
incident wave packet (\ref{fp}) is decomposed into two independent
constituents oppositely polarized along $x$-direction and moving
with different velocities. The spatial distribution of the
electron density is given by the integral over all momenta, i.e.
\begin{eqnarray}
\label{split} \rho_\pm (y,t) = \frac{1}{2\pi^{1/2} \delta x (t)}
\exp{[-\frac{(x-\bar v_\pm t)^2}{\delta x^2 (t)}]},
\end{eqnarray}
with the average velocities $\bar v_\pm =\bar p/m \pm \alpha$, and
the Gaussian width at finite times, $\delta x^2 (t) =\delta
p^{-2}+\frac{t^2\delta p^2}{m^2}$. To observe the spin-orbit
induced splitting of a wave packet the following conditions should
be satisfied,
$$
\frac{\delta p}{m} \ll \alpha \ll \frac{t}{\delta p}.
$$
The first of the two conditions ensures that the splitting
dominates over the wave packet broadening, while the second
condition means that enough time has to elapse before the
splitting becomes larger than the intrinsic packet width.

Now let us consider an injection of a packet initially polarized
along the $y$-direction.
\begin{equation}
\label{fp2} \hat f_{\bf p}({\bf x}) = (\hat \sigma_0 +\hat
\sigma_y) F (y),
\end{equation}
The equations for the $f$ and $g_x$ components of the density
matrix remain unchanged with the above analysis still valid. The
second pair of Eqs. (\ref{packet}) is independent of the first
pair and have a solution,
\begin{eqnarray}
\label{precession}
g_y(y,t) = F(y-vt)\cos{(2\Delta_pt)},\nonumber\\
g_z(y,t) = -F(y-vt)\sin{(2\Delta_pt)}.
\end{eqnarray}
According to the expressions (\ref{precession}) the initial spin
polarization precesses with a frequency $2\Delta_p$ around the
axis perpendicular to the propagation direction. Note that the
precessing spin propagates with the center-of-mass velocity $\bar
v$ rather than with the subband velocities $\bar v_\pm$.

The above analysis assumes that a wavepacket is injected  with a
given momentum $\bar p$. Such an injection into 2DEG with a
spin-orbit coupling is not easy to achieve. For example, injection
through an interface with a 'normal' (with no spin-orbit
interaction) 2DEG\cite{HM,MH,LLF} would not result in a spatial
splitting of a wave packet. This is due to the fact that the
injection happens with a conservation of energy rather than
momentum. As seen from Eqs. (\ref{spectrum}-\ref{spectrum1}) the
two states with the same energy propagate with the same
velocity\cite{MSB,LLF} within the approximations of this paper.
However, if we take into account the cubic Dresselhaus terms,
which have been omitted in our discussion, there can be a
splitting of velocities at the same energy. In order to achieve
splitting without the cubic terms we need to consider a more
complicated setup. As a demonstration of principle, we consider
the following example. Let us inject a wave packet propagating
along the $y$-direction with the spin polarized along the
interface ($x$-axis), e.g. by injection from a ferromagnetic
contact. The states forming the wavepacket belong to the subband
1, with a spin polarization $\frac{1}{\sqrt 2} (1,1)$. Let us now
switch on ac magnetic field along the $y$-axis rotating the spin
direction until it is aligned with the $z$-axis, $(1,0)$, and then
switch the magnetic field off. The resulting state will be an
equal mixture of both eigenstates $\frac{1}{\sqrt 2} (1,1)$ and
$\frac{1}{\sqrt 2} (1,-1)$  without any change of momentum (the
energy is no longer conserved). The velocities of these states are
different and the packet will split.

The above picture holds not only for the injection of initially
polarized packet. If the incident packet is unpolarized and has a
given energy, upon entering the interface it will become a mixture
of two states: $\frac{1}{\sqrt 2} (1,1)$ with the momentum
$p_0-m\alpha$, and $\frac{1}{\sqrt 2} (1,-1)$ with the momentum
$p_0+m\alpha$. Both velocities remain equal to $v_0=p_0/m$. After
switching on the ac magnetic field with the frequency $\omega
\approx 2\alpha p_0$ (which is a resonant frequency for the
transition between the two subbands), the first state will evolve
into the mixture of the states: $\frac{1}{\sqrt 2} (1,1)$ and
$\frac{1}{\sqrt 2} (1,-1)$, both with the momentum $p_0+m\alpha$,
meaning two different velocities $v_0$ and $v_0-2\alpha$. The same
reasoning shows that the other initial state will develop two
velocities $v_0$ and $v_0+2\alpha$. Therefore, the initially
unpolarized packet will split into three parts.

{\it Ballistic spin injection}. We envisage a spin injection from
ferromagnetic contacts into ballistic 2DEG  among the applications
for the equations derived above. In this case the injection occurs
with  conservation of energy, and can be described by the
time-independent solution of the equations (\ref{tr1}-\ref{tr2})
with the appropriate boundary conditions, which require a
conservation of the normal components of the electric current,
Eq.\ (\ref{density}), at the interfaces. A corresponding theory
would generalize the existing approach for the ballistic
spin-injection based on the ordinary Boltzmann
equation\cite{Rashba}. In the latter case the Boltzmann equation
method is more convenient for the calculation of spin polarization
of current and magnetoresistance than the direct solution of the
single-particle Schr\"odinger equation.

 {\it Structure factor.} The electron density
fluctuations are described by the structure factor \cite{PN}
defined as the retarded correlation function,
\begin{eqnarray}
\label{struc} \chi(x,x') = -i\theta(t-t') \langle \langle
\rho(x)\rho (x')-\rho (x') \rho (x) \rangle \rangle,
\end{eqnarray}
of the electron density operators $\rho (x)= \psi_\alpha^\dagger
(x) \psi_\alpha (x)$. At equilibrium the structure factor
(\ref{struc}) depends on the relative coordinates $x-x'$ only. The
imaginary part of the Fourier transform $\chi (\omega,{\bf q})$
measures the energy dissipation of the external field at a given
frequency $\omega$ and a wavevector ${\bf q}$. In the isotropic
system the structure factor is related to the ac conductivity by
the relation,
\begin{eqnarray} \label{conduc} \sigma (\omega)
=\lim_{q \to 0} \frac{ie^2\omega}{q^2} \chi (\omega,{\bf q}).
\end{eqnarray}
The formula (\ref{conduc}) is readily checked using the Kubo
formula for the conductivity and the continuity equation
(\ref{cont}).

 According to the fluctuation-dissipation theorem, the structure
 factor can be determined from calculations of the linear response
 to an
external scalar field. The field-induced modulation of electron
density is related to the magnitude of the external perturbation
through the structure factor according to \cite{PN},
\begin{eqnarray}
\label{relation} \delta \rho (\omega,{\bf q}) = e \chi(\omega,{\bf
q}) \phi (\omega,{\bf q}).
\end{eqnarray}
The electron density modulation is given by the deviation of the
function $f_{\bf p}(t,{\bf x})$ from its equilibrium value,
$\delta \rho (\omega,{\bf q }) =  \int d{\bf p} ~\delta f_{\bf p}
(\omega,{\bf q }),$ and can be found from the linearized equations
(\ref{tr1},\ref{tr2}).
  In the linear approximation by the external
field $\phi(\omega, {\bf q})$, the distribution function is a
small deviation
\begin{eqnarray}
\label{delta}
 f_{\bf p} =f^0_{\bf p} +\delta f_{\bf p},~~~
{\bf g}_{\bf p} = {\bf g}^0_{\bf p}+\delta {\bf g}_{\bf p},
\end{eqnarray}
from its equilibrium value (\ref{equil}). The linearized transport
equations (\ref{tr1},\ref{tr2}) take the form,
\begin{eqnarray}
\label{linear} (\omega -{\bf qv} )\delta f_{\bf p}&-&\alpha_{ik}
q_i \delta {g}_{{\bf p}k}= e\phi (\omega, {\bf
q}) (f^0_{{\bf p}+\frac{\bf q}{2}}+f^0_{{\bf p}-\frac{\bf q}{2}}),\nonumber \\
(\omega -{\bf qv} ) \delta {g}_{{\bf p}i}&-&i [{\bf b}_{\bf p}
\times \delta {\bf g}_{\bf p}]_i -\alpha_{ik} q_k \delta f_{\bf p} = \nonumber\\
 &-& e \phi (\omega, {\bf
q}) ({g}^0_{{\bf p}+\frac{\bf q}{2}i}+{\bf g}^0_{{\bf p}-\frac{\bf
q}{2}i}).
\end{eqnarray}
Solving these equations for the variation of the electron density
(\ref{density}) we obtain  the structure factor with the help of
the relation (\ref{relation}),
\begin{eqnarray}
\label{chi} \chi(\omega, {\bf q}) &=& \frac{1}{2} \sum_{\mu \mu'}
\int d{\bf p} ~  [1+(-1)^{\mu\mu'}\cos{(\chi_{{\bf p}}-\chi_{{\bf
p'}})}] \nonumber\\ && \times \frac{n_{\mu {\bf p}_-}-n_{\mu'{\bf
p}_+}}{\omega-\epsilon_{\mu'{\bf p}_+}+\epsilon_{\mu {\bf p}_-}},
\end{eqnarray}
 where ${\bf p}_\pm={\bf p}\pm {\bf q}/2$. The expression
 (\ref{chi}) with $\omega=0$ corresponds to the previously derived result
  for the static dielectric
 function\cite{CR}.
 To simplify further the subsequent discussion we
will disregard the anisotropy, $\beta =0$, and consider the
zero-temperature limit $T=0$. The two spin-orbit subbands are
axially symmetric, shown on Fig. 1. The subbands are filled up to
the same Fermi energy level $\epsilon_F$ but have two different
Fermi momenta, $p_1$ and $p_2$, determined from the equations
$\epsilon_i(p_i)= \epsilon_F$,  where $\epsilon_i(p)$ are given by
Eqs. (\ref{spectrum},\ref{spectrum1}) with $\beta =0$. This leads
to the values,
\begin{eqnarray}
p_1= p_0-m\alpha +O(m^2\alpha^2/p_0^2),\nonumber\\
p_2= p_0+m\alpha +O(m^2\alpha^2/p_0^2),
\end{eqnarray}
where $p_0$ is determined by $\epsilon_F=p_0^2/2m$, namely $p_0$
is the Fermi-momentum in the absence of spin-orbit interaction.
Note that the Fermi velocities for the two subbands,
\begin{equation}
v_i = \frac{\partial \epsilon_i(p)}{\partial p} \vert_{p=p_i} =
\frac{p_0}{m}+O(m^2\alpha^2/p_0^2),
\end{equation}
are the same and (up to higher order terms) equal to the Fermi
velocity in 2DEG with no spin-orbit coupling $\alpha =0$.
\begin{figure}
\resizebox{.33\textwidth}{!}{\includegraphics{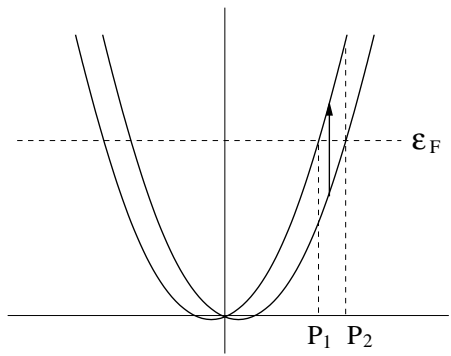}}
\caption{Spin-orbit induced subbands of an isotropic
two-dimensional electron gas, $\epsilon_p=p^2/2m \pm \alpha |p|$.
At $T=0$ all states below the Fermi energy $\epsilon_F$ are
filled. The Fermi momenta for the two subbands are $p_{1,2}=p_0\mp
m\alpha$. The direct transitions, $q=0$ (shown by the arrow) are
possible for the states between the dashed lines, $p_1 < p <p_2$.}
\end{figure}
The imaginary part of the structure factor $\chi(\omega,{\bf q})$
determines the absorption, or Landau damping, of the external
field  at given frequency and wavevector. The points in the
electron momentum space  that contribute to the Landau damping
correspond to the zeros of the denominators. There are total four
determined by the equations,
\begin{equation}
\label{excit} \omega ={\bf qv} \pm \alpha {p}_+\pm \alpha  p_-,
\end{equation}
with the opposite signs of the last two terms corresponding to the
(gapless) transitions within the same subbands [Eq.\
(\ref{spectrum})] and equal signs describing the transitions
between different subbands.

 The terms with $\mu=\mu'$ in Eq. (\ref{chi}) represent the effect of
intrasubband transitions. Only indirect ($q \ne 0$) transitions
contribute to the imaginary part of the structure factor. For
small transferred momenta $q \ll p_0$ one can disregard the
deviation of the cosine factor from unity and also approximate
$n_{i{\bf p}_-}-n_{i{\bf p}_+} \simeq - {\bf q} \partial n_{i}
/\partial {\bf p} $. Taking the momentum integral we obtain for
the contribution of the $i$-th subband
\begin{eqnarray}
\label{Im} \Im \chi_i (\omega,q)&=& -\nu_i \frac{
\omega}{\sqrt{q^2v_0^2 -\omega^2 }}~\theta(q^2v^2_0-\omega^2),
\end{eqnarray}
where $v_0=p_0/m$ and $\nu_i$ stands for the density of states of
the $i$-th subband at its Fermi surface $p=p_i$:
$\nu_1=\frac{m}{2\pi} (1-\frac{m\alpha}{p_0})$,
$\nu_2=\frac{m}{2\pi} (1+\frac{m\alpha}{p_0})$. Note that the sum
of the two contributions (\ref{Im}) is {\it independent} of the
spin-orbit interaction (up to higher-order terms), a consequence
of the fact that the two subbands have the same value of the
Fermi-velocity. Spin-orbit interaction results only in a
redistribution of the spectral weight between the subbands
controlled by the changes in the densities of states.

The terms with $\mu\ne \mu'$ in Eq. (\ref{chi}) correspond to the
intersubband transitions. Their contribution to the structure
factor for $m\alpha \ll q \ll p_0$ is negligible compared to the
above considered intrasubband transitions by the factor $\sim
q^2/p_0^2$ (due to the small $\sin^2$ prefactor). However, the
presence of the two subbands is important as it makes the direct,
$q=0$, transitions possible. The factor $n_{1p}-n_{2p}$ then
defines the momentum space available for the direct transitions, $
p_1 < p <p_2$ (see Fig. 1), which corresponds to the frequency
domain $2\Delta_0 - 2m\alpha^2 <\omega < 2\Delta_0 +2m\alpha^2$,
where $\Delta_0= \Delta_{p_0}$,
\begin{equation}
\label{direct} \chi (\omega,q \to 0) = \frac{\alpha q^2}{4\pi}
\int\limits_{p_1}^{p_2} \frac{dp}{(\omega+i0)^2-4\Delta_p^2}.
\end{equation}
 The imaginary part of this expression is
\begin{equation}
\label{imdirect} \Im \chi(\omega,q \to 0) = -\frac{q^2 \text{sgn}~
\omega }{32\Delta_0} ~ \theta [4m^2\alpha^4 -(\omega-
2\Delta_0)^2].
\end{equation}
The equation (\ref{direct}) corresponds to the previously obtained
result \cite{MCE} for the optical conductivity $\sigma (\omega)$.
The expression (\ref{direct}) goes to zero with the wavevector,
which is easily understood by noting that the matrix elements for
the transitions between $\psi_1({\bf p}_-) $ and $\psi_{2}({\bf p
}_+)$ states are suppressed at small transferred momenta since
they are orthogonal at $q=0$. However, their contribution to the
conductivity [according to Eq. (\ref{conduc})] remains finite,
which is clear since the operator of electron velocity has nonzero
matrix elements for the intersubband transitions even at $q=0$.

The experimental observation of the direct transitions
(\ref{imdirect}) is feasible in the measurements of the resonant
microwave absorption in high-mobility semiconductor
heterostructures.

 {\it Screened electron-electron interaction and plasmon
excitations.} So far our analysis has been restricted to the
noninteracting electron gas. To incorporate the effects of the
electron-electron interaction  in the random phase approximation
one has to account for the self-consistent  electric field
 induced by the variations of the electron density. The potential for this field
 $\phi_{sc}$ obeys the Poisson equation. In two dimensions the Fourier
transform of the Poisson equation has the form,
\begin{equation}
\label{poisson} e\phi_{sc} (\omega,{\bf q})= V_q \rho (\omega,{\bf
q}),
\end{equation}
where $V_q=2\pi e^2/q$ is the bare Coulomb propagator. The random
phase approximation (RPA) is then equivalent to the substitution
$\phi (\omega,{\bf q}) \to \phi_{sc} (\omega,{\bf q})+ \phi
(\omega,{\bf q})$ in the right-hand side of Eq. (\ref{linear}). It
is straightforward to see that the structure factor takes the
familiar RPA form,
\begin{equation}
\label{RPA} \chi_{RPA}(\omega,{\bf q})=\frac{\chi (\omega,{\bf
q})}{1-V_q \chi (\omega,{\bf q})}.
\end{equation}
The pole of this expression determines the plasmon spectrum
$\omega=\omega_q +i\gamma_q$, where $\omega_q^2 = v^2 \kappa q/2$,
with $\kappa=2\pi e^2 \nu$ standing for the static screening
radius. The plasmon linewidth is given by the imaginary part of
the bare structure factor at $\omega =\omega_q$,
\begin{equation}
\label{damping} \gamma_q = \frac{1}{2} V_q\omega_q |\Im
\chi(\omega_q,{q})|.
\end{equation}
For the plasmon to be an undamped excitation its frequency should
lie above the electron-hole continuum, $\omega_q > qv$, which
requires $\kappa
> 2q$. As was already pointed out in Ref. \onlinecite{MCE}
the plasmon acquires damping when $\omega_q \sim 2\Delta$. Since
$q \ll (q \kappa)^{1/2} \sim m\alpha$ at this range, the direct
transitions (\ref{imdirect}) make  the principal contribution to
Eq.\ (\ref{damping}).

{\it Impurity scattering}. The equations presented in this paper
assume ballistic electron motion. The absence of impurities allows
one to write kinetic equation as a closed set of equations, Eq.\
(\ref{tr}), for the density matrix integrated over the energy
variable $\epsilon$, [see Eq.\ (\ref{pauli})], i.e.\ at coinciding
times. In the presence of disorder the self-energy due to impurity
scattering should be added to the right-hand side of Eq.\
(\ref{set}). In general, since plain waves are no longer
eigenstates of the system with impurities, the equations for the
distribution function depending on the momentum ${\bf p}$ (and not
on the energy $\epsilon$) become not very convenient. More natural
(though more complicated) equations would result from integration
over $\xi_p$, similar to the usual spin-degenerate case\cite{RS}.
Such equations are beyond the scope of the present paper.

{\it Special case $\alpha=\pm \beta$.} Recently, Schliemann et.\
al.\ \cite{SEL} proposed a spin field-effect transistor based on a
particular tuning of the spin-orbit coupling constants such that
$\alpha=\beta$ (or $\alpha=-\beta$). This special system is
expected to preserve spin coherence even in the presence of
disorder. This is due to the fact that the spin eigenstates
(\ref{states}) are independent of the electron momenta, $\chi_{\bf
p} = \text{const}$, therefore a scalar impurity potential does not
result in the intersubband transitions.  The same observation
holds for the structure factor. Since the matrix elements of the
density are identically zero for the transitions between different
subbands the second line of Eq. (\ref{chi}) is absent in this case
and the structure factor is intact by the presence of spin-orbit
interaction (up to higher order corrections).

{\it Conclusions.} To summarize, we have derived transport
equations for the distribution function of a two-dimensional
electron gas with spin-orbit interaction of both the
Bychkov-Rashba and the Dresselhaus mechanisms. The distribution
function is a 2$\times$2-matrix in the spin space. General
expressions for the particle and energy currents and densities are
available in terms of the density $f_{\bf p}$ and spin ${\bf
g}_{\bf p}$ distribution functions. The obtained equations are
applied to the wave-packet propagation in a ballistic 2DEG and to
the calculation of the density-density correlation function
$\chi(\omega,q)$. We observe that for $q > m\alpha$ the structure
factor $\chi(\omega,q)$ is almost not affected by the spin-orbit
interaction, but it reveals new features when $q \ll m\alpha$ due
to the direct transitions between different spin-orbit subbands.

 Fruitful discussions with A.\ Andreev, K.\ Flensberg, L.
Glazman, and C.\ Marcus are gratefully appreciated. This material
is based on work supported by the NSF under grant PHY-01-17795 and
by the Defence Advanced Research Programs Agency (DARPA) under
Award No. MDA972-01-1-0024.

\end{document}